\documentclass{svmult} 
\usepackage{amsmath}
\usepackage{amssymb} 
\usepackage{amscd}

\def\Tor{{\mbox{Tor}}}

\def\dim{{\mbox{dim}}}
\def\ker{{\mbox{Ker}}}     
\def\Der{{\mbox{Der}}}
\def\Hom {{\mbox{Hom}}}

\def\Int{{\mbox{Int}}}

\def\cala{{\cal A}} 
\def\calb{{\cal B}} 
\def\calk{{\cal K}} 
\def\call{{\cal L}}
\def\calm{{\cal M}}

\def\cals{{\cal S}}

\def\fraca{{\mathfrak A}}

\def\bbbone{\mbox{\rm 1\hspace {-.6em} l}}

\def\gg{{\mathbf g}}

\def\Lg{{\mathbf L}}
\def\Rg{{\mathbf R}}

\def\zig{{\mathbf \zeta}}

\begin{document}

 \title{Yang-Mills and some related algebras}

\author{Alain CONNES  and Michel DUBOIS-VIOLETTE}
\institute{Coll\`ege de France, 3 rue d'Ulm, 75  005 Paris,  I.H.E.S. and Vanderbilt University, connes$@$ihes.fr \\
Laboratoire de Physique Th\'eorique, UMR 8627, Universit\'e Paris XI,
B\^atiment 210, F-91 405 Orsay Cedex, France, 
Michel.Dubois-Violette$@$th.u-psud.fr}

\maketitle
\begin{center}
{\sl Dedicated to Jacques Bros}
\end{center}
\begin{abstract}
After a short introduction on the theory of homogeneous algebras we describe the application of this theory to the analysis of the cubic Yang-Mills algebra, the quadratic self-duality algebras, their ``super" versions as well as to some generalization.
\end{abstract}

\noindent LPT-ORSAY 04-108

\section{Introduction}
Consider the classical Yang-Mills equations in $(s+1)$-dimensional pseudo Euclidean space $\mathbb R^{s+1}$ with pseudo metric denoted by $g_{\mu\nu}$ in the canonical basis of $\mathbb R^{s+1}$ corresponding to coordinates $x^\lambda$. For the moment the signature plays no role so $g_{\mu\nu}$ is simply a real nondegenerate symmetric matrix with inverse denoted by $g^{\mu\nu}$. In terms of the covariant derivatives $\nabla_\mu=\partial_\mu + A_\mu$ ($\partial_\mu=\partial / \partial x^\mu)$ the Yang-Mills equations read
\begin{equation}
g^{\lambda \mu} [\nabla_\lambda, [\nabla_\mu,\nabla_\nu]]=0
\label {YM}
\end{equation}
for $\nu\in \{0,\dots,s\}$. By forgetting the detailed origin of these equations, it is natural to consider the abstract unital associative algebra $\cala$ generated by $(s+1)$ elements $\nabla_\lambda$ with the $(s+1)$ cubic relations (\ref{YM}). This algebra will be refered to as the Yang-Mills algebra. It is worth noticing here that Equations (\ref{YM}) only involve the product through commutators so that, by its very definition the Yang-Mills algebra $\cala$ is a universal enveloping algebra.

Our aim here is to present the analysis of the Yang-Mills algebra and of some related algebras based on the recent development of the theory of homogeneous algebras \cite{ber:2001a}, \cite{ber-mdv-wam:2003}. This analysis is only partly published in \cite{ac-mdv:2002b}.\\

In the next section we recall some basic concepts and results on homogeneous algebras which  will be used in this paper.\\

Section 3 is devoted to the Yang-Mills algebra. In this section we recall the definitions and the results of 
\cite{ac-mdv:2002b}. The proofs are omitted since these are in \cite{ac-mdv:2002b} and since very similar proofs are given in Sections 4 and 6. Instead, we describe the structure of the bimodule resolution of the Yang-Mills algebra and the structure of the corresponding small bicomplexes which compute the Hochschild homology.\\

In Section 4 we define the super Yang-Mills algebra and we prove for this algebra results which are the counterpart of the results of \cite{ac-mdv:2002b} for the Yang-Mills algebra.\\

In Section 5 we define and study the super self-duality algebra. In particular, we prove for this algebra the analog of the results of \cite{ac-mdv:2002b} for the self-duality algebra and we point out a very surprising connection between the super self-duality algebra and the algebras occurring in our analysis of noncommutative 3-spheres 
\cite{ac-mdv:2002a}, \cite{ac-mdv:2003}.\\

In Section 6 we describe some deformations of the Yang-Mills algebra and of the super Yang-Mills algebra.

\section{Homogeneous algebras}

 Although we shall be concerned in the following with the cubic Yang-Mills algebra $\cala$,  the quadratic self-duality algebra $\cala^{(+)}$ \cite{ac-mdv:2002b} and some related algebras, we recall in this section some constructions and some results for general $N$-homogeneous algebras  \cite{ber-mdv-wam:2003}, \cite{ber:2001a}. All vector spaces are over a fixed commutative field $\mathbb K$.\\

A {\sl homogeneous algebra of degree $N$ or $N$-homogeneous algebra} is an algebra of the form 
\[
\cala = A(E,R)=T(E)/(R)
\]
where $E$ is a finite-dimensional vector space, $R$ is a linear subspace of $E^{\otimes^N}$ and where $(R)$ denotes the two-sided ideal of the tensor algebra $T(E)$ of $E$ generated by $R$. The algebra $\cala$ is naturally a connected graded algebra with graduation induced by the one of $T(E)$. To $\cala$ is associated another $N$-homogeneous algebra, {\sl its dual}  $\cala^!=A(E^\ast, R^\perp)$ with $E^\ast$ denoting the dual vector space of $E$ and $R^\perp\subset E^{\otimes^N\ast}=E^{\ast \otimes^N}$ being the annihilator of $R$, \cite{ber-mdv-wam:2003}. The $N$-complex $K(\cala)$ of left $\cala$-modules is then defined to be
\begin{equation}
\dots \stackrel{d}{\rightarrow} \cala\otimes \cala^{!\ast}_{n+1} \stackrel{d}{\rightarrow} \cala\otimes \cala^{!\ast}_{n}\stackrel{d}{\rightarrow} \dots \stackrel{d}{\rightarrow}\cala \rightarrow 0
\label{eq2.1}
\end{equation}
where $\cala^{!\ast}_n$ is the dual vector space of the finite-dimensional vector space $\cala^!_n$ of the elements of degree $n$ of $\cala^!$ and where $d:\cala\otimes \cala^{!\ast}_{n+1}\rightarrow \cala\otimes \cala^{!\ast}_n$ is induced by the map $a\otimes (e_1\otimes \dots \otimes e_{n+1})\mapsto ae_1 \otimes (e_2\otimes \dots \otimes e_{n+1})$ of $\cala\otimes E^{\otimes^{n+1}}$ into $\cala\otimes E^{\otimes^n}$, remembering that $\cala^{!\ast}_n\subset E^{\otimes^n}$, (see \cite{ber-mdv-wam:2003}). In (\ref{eq2.1}) the factors $\cala$ are considered as left $\cala$-modules. By considering $\cala$ as right $\cala$-module and by exchanging the factors one obtains the $N$-complex $\tilde K(\cala)$ of right $\cala$-modules
\begin{equation}
\dots \stackrel{\tilde d}{\rightarrow} \cala^{!\ast}_{n+1} \otimes \cala \stackrel{\tilde d}{\rightarrow} \cala^{!\ast}_n\otimes \cala \stackrel{\tilde d}{\rightarrow} \dots \stackrel{\tilde d}{\rightarrow} \cala \rightarrow 0
\label{eq2.2}
\end{equation}
where now $\tilde d$ is induced by $(e_1\otimes \dots \otimes e_{n+1})\otimes a \mapsto (e_1\otimes \dots \otimes e_n)\otimes e_{n+1}a$. Finally one defines two $N$-differentials $d_{\Lg}$ and $d_{\Rg}$ on the sequence of $(\cala,\cala)$-bimodules, i.e. of left $\cala\otimes \cala^{opp}$-modules, $(\cala\otimes \cala^{!\ast}_n \otimes \cala)_{n\geq 0}$ by setting $d_\Lg =d\otimes I_\cala$ and $d_\Rg = I_\cala\otimes \tilde d$ where $I_\cala$ is the identity mapping of $\cala$ onto itself. For each of these $N$-differentials $d_\Lg$ and $d_\Rg$ the sequences
\begin{equation}
\dots \stackrel{d_\Lg,d_\Rg}{\rightarrow} \cala \otimes \cala^{!\ast}_{n+1}\otimes \cala \stackrel{d_\Lg,d_\Rg}{\rightarrow} \cala\otimes \cala^{!\ast}_n \otimes \cala \stackrel{d_\Lg, d_\Rg}{\rightarrow}\dots
\label{eq2.3}
\end{equation}
are $N$-complexes of left $\cala\otimes \cala^{opp}$-modules and one has
\begin{equation}
d_\Lg d_\Rg = d_\Rg d_\Lg
\label{eq2.4}
\end{equation}
which implies that
\begin{equation}
d^N_\Lg -d^N_\Rg = (d_\Lg -d_\Rg)\left ( \sum^{N-1}_{p=0} d^p_\Lg d^{N-p-1}_\Rg\right) = \left( \sum^{N-1}_{p=0} d^p_\Lg d^{N-p-1}_\Rg\right) (d_\Lg -d_\Rg) =0
\label{eq2.5}
\end{equation}
in view of $d^N_\Lg=d^N_\Rg=0$.\\

As for any $N$-complex \cite{mdv:1998a} one obtains from $K(\cala)$ ordinary complexes $C_{p,r}(K(\cala))$, {\sl the contractions of} $K(\cala)$, by putting together alternatively $p$ and $N-p$ arrows $d$ of $K(\cala)$. Explicitely $C_{p,r}(K(\cala))$ is given by 
\begin{equation}
\dots \stackrel{d^{N-p}}{\rightarrow} \cala \otimes \cala^{!\ast}_{Nk+r}\stackrel{d^p}{\rightarrow} \cala\otimes \cala^{!\ast}_{Nk-p+r}\stackrel{d^{N-p}}{\rightarrow}\cala\otimes \cala^{!\ast}_{N(k-1)+r}\stackrel{d^p}{\rightarrow}\dots
\label{eq2.6}
\end{equation}
for $0\leq r< p\leq N-1$, \cite{ber-mdv-wam:2003} . These are here chain complexes of free left $\cala$-modules. As shown in \cite{ber-mdv-wam:2003} the complex $C_{N-1,0}(K(\cala))$ coincides with the {\sl Koszul complex} of \cite{ber:2001a}; this complex will be denoted by $\calk(\cala,\mathbb K)$ in the sequel. That is one has
\begin{equation}
\calk_{2m}(\cala,\mathbb K)=\cala\otimes\cala^{!\ast}_{Nm},\ \ \ \calk_{2m+1}(\cala,\mathbb K)=\cala\otimes \cala^{!\ast}_{Nm+1}
\label{eq2.7}
\end{equation}
for $m\geq 0$, and the differential is $d^{N-1}$ on $\calk_{2m}(\cala,\mathbb K)$ and $d$ on $\calk_{2m+1}(\cala,\mathbb K)$. If $\calk(\cala, \mathbb K)$ is acyclic in positive degrees then $\cala$ will be said to be a {\sl Koszul algebra}. It was shown in \cite{ber:2001a} and this was confirmed by the analysis of \cite{ber-mdv-wam:2003} that this is the right generalization for $N$-homogeneous algebra of the usual notion of Koszulity for quadratic algebras  \cite{man:1987}, \cite{lod:1999}. One always has $H_0(\calk(\cala,\mathbb K))\simeq \mathbb K$ and therefore if $\cala$ is Koszul, then one has a free resolution $\calk(\cala,\mathbb K)\rightarrow \mathbb K \rightarrow 0$ of the trivial left $\cala$-module $\mathbb K$, that is the exact sequence
\begin{equation}
\dots \stackrel{d^{N-1}}{\rightarrow} \cala \otimes \cala^{!\ast}_{N+1}\stackrel{d}{\rightarrow} \calaÊ\otimes R \stackrel{d^{N-1}}{\rightarrow}\cala\otimes E\stackrel{d}{\rightarrow} \cala \stackrel{\varepsilon}{\rightarrow} \mathbb K \rightarrow 0
\label{eq2.8}
\end{equation}
of left $\cala$-modules where $\varepsilon$ is the projection on degree zero. This resolution is a minimal projective resolution of $\cala$  in the graded category  \cite{ber:2002a}.\\

One defines now the chain complex of free $\cala\otimes\cala^{opp}$-modules $\calk(\cala,\cala)$ by setting
\begin{equation}
\calk_{2m}(\cala,\cala)=\cala\otimes \cala^{!\ast}_{Nm}\otimes \cala,\ \ \ \calk_{2m+1}(\cala,\cala)=\cala\otimes \cala^{!\ast}_{Nm+1}\otimes \cala
\label{eq2.9}
\end{equation}
for $m\in \mathbb N$ with differential $\delta'$ defined by
\begin{equation}
\delta'=d_\Lg -d_\Rg :\calk_{2m+1}(\cala,\cala)\rightarrow \calk_{2m}(\cala,\cala)
\label{eq2.10}
\end{equation}
\begin{equation}
\delta'=\sum^{N-1}_{p=0} d^p_\Lg d^{N-p-1}_\Rg:\calk_{2(m+1)}(\cala,\cala)\rightarrow \calk_{2m+1}(\cala,\cala)
\label{eq2.11}
\end{equation}
the property $\delta^{\prime 2}=0$ following from (\ref{eq2.5}). {\sl This complex is acyclic in positive degrees if and only if $\cala$ is Koszul}, that is if and only if $\calk(\cala,\mathbb K)$ is acyclic in positive degrees,  \cite{ber:2001a} and \cite{ber-mdv-wam:2003} . One always has the obvious exact sequence
\begin{equation}
\cala\otimes E\otimes \cala \stackrel{\delta'}{\rightarrow} \cala\otimes \cala \stackrel{\mu}{\rightarrow}\cala\rightarrow 0
\label{eq2.12}
\end{equation}
of left $\cala\otimes \cala^{opp}$-modules where $\mu$ denotes the product of $\cala$. It follows that if $\cala$ is a Koszul algebra then $\calk(\cala,\cala)\stackrel{\mu}{\rightarrow}\cala\rightarrow 0$ is a free resolution of the $\cala\otimes \cala^{opp}$-module $\cala$ which will be refered to as {\sl the Koszul resolution of} $\cala$. This is a minimal projective resolution of $\cala\otimes \cala^{opp}$ in the graded category \cite{ber:2002a}.\\

Let $\cala$ be a Koszul algebra and let $\calm$ be a $(\cala,\cala)$-bimodule considered as a right $\cala\otimes \cala^{opp}$-module. Then, by interpreting the $\calm$-valued Hochschild homology $H(\cala,\calm)$ as $H_n(\cala,\calm)=\Tor_n^{\cala\otimes \cala^{opp}}(\calm,\cala)$  {\cite{car-eil:1973},  the complex $\calm \otimes_{\cala\otimes \cala^{opp}}\calk(\cala,\cala)$ computes the $\calm$-valued Hochschild homology of $\cala$, (i.e. its homology is the ordinary $\calm$-valued Hochschild homology of $\cala$). We shall refer to this complex as {\sl the small Hochschild complex of} $\cala$ with coefficients in $\calm$ and denote it by $\cals(\cala, \calm)$. It reads
\begin{equation}
\dots \stackrel{\delta}{\rightarrow}\calm\otimes \cala^{!\ast}_{N(m+1)}\stackrel{\delta}{\rightarrow} \calm \otimes \cala^{!\ast}_{Nm+1}\stackrel{\delta}{\rightarrow}\calm\otimes \cala^{!\ast}_{Nm} \stackrel{\delta}{\rightarrow}\dots
\label{eq2.13}
\end{equation}
where $\delta$ is obtained from $\delta'$ by applying the factors $d_L$ to the right of $\calm$ and the factors $d_R$ to the left of $\calm$.\\

Assume that $\cala$ is a Koszul algebra of finite global dimension $D$. Then the Koszul resolution of $\mathbb K$ has length $D$, i.e. $D$ is the largest integer such that $\calk_D(\cala,\mathbb K)\not=0$. By construction, $D$ is also the greatest integer such that $\calk_D(\cala,\cala)\not= 0$ so the free $\cala\otimes \cala^{opp}$-module resolution of $\cala$ has also length $D$. Thus {\sl for a Koszul algebra, the global dimension is equal to the Hochschild dimension}. Applying then the functor $\Hom_\cala(\bullet, \cala)$ to $\calk(\cala,\mathbb K)$ one obtains the cochain complex $\call(\cala,\mathbb K)$ of free right $\cala$-modules
\[
0\rightarrow \call^0(\cala,\mathbb K)\rightarrow \dots \rightarrow \call^D(\cala,\mathbb K) \rightarrow 0
\]
where $\call^n(\cala,\mathbb K)=\Hom_\cala(\calk_n(\cala,\mathbb K),\cala)$. The Koszul algebra $\cala$ is {\sl Gorenstein} iff $H^n(\call(\cala,\mathbb K))=0$ for $n<D$ and $H^D(\call(\cala,\mathbb K))=\mathbb K$ (= the trivial right $\cala$-module). This is clearly a generalisation of the classical Poincar\'e duality and this implies a precise form of Poincar\'e duality between Hochschild homology and Hochschild cohomology \cite{ber-mar:2003}, \cite{vdb:1998}, \cite{vdb:2002}. In the case of the Yang-Mills algebra and its deformations which are Koszul Gorenstein cubic algebras of global dimension 3, this Poincar\'e duality gives isomorphisms
\begin{equation}
H_k(\cala,\calm)= H^{3-k}(\cala,\calm),\>\>\> k\in \{0,1,2,3\}
\label{Pd}
\end{equation}
between the Hochschild homology and the Hochschild cohomology with coefficients in a bimodule $\calm$.

\section{The Yang-Mills algebra}

Let $(g_{\lambda\mu})\in M_{s+1}(\mathbb K)$ be an invertible symmetric $(s+1)\times (s+1)$-matrix with inverse $(g^{\lambda\mu})$, i.e. $g_{\lambda\mu} g^{\mu\nu}=\delta^\nu_\lambda$. The {\sl Yang-Mills algebra} is the cubic algebra $\cala$ generated by $s+1$ elements $\nabla_\lambda$ $(\lambda\in \{0,\dots,s\})$ with the $s+1$ relations
\[
g^{\lambda\mu}[\nabla_\lambda,[\nabla_\mu,\nabla_\nu]]=0,\>\>\> \nu\in \{0,\dots,s\}
\]
that is $\cala=A(E,R)$ with $E=\oplus_\lambda \mathbb K \nabla_\lambda$ and $R\subset E^{\otimes^3}$ given by
\begin{eqnarray}
R & = &\sum_\nu \mathbb K g^{\lambda\mu}[\nabla_\lambda,[\nabla_\mu,\nabla_\nu]_\otimes]_\otimes\nonumber\\
& = & \sum_\rho \mathbb K(g^{\rho\lambda}g^{\mu\nu}+g^{\nu\rho}g^{\lambda\mu}-2g^{\rho\mu}g^{\lambda \nu})\nabla_\lambda \otimes \nabla_\mu \otimes \nabla_\nu
\label{rYM}
\end{eqnarray}
In \cite{ac-mdv:2002b} the following theorem was proved.
\begin{theorem}
The cubic Yang-Mills algebra $\cala$ is Koszul of global dimension 3 and is Gorenstein.
\end{theorem}
The proof of this theorem relies on the computation of the dual cubic algebra $\cala^!$ which we now recall.\\

The dual $\cala^!=A(E^\ast,R^\perp)$ of the Yang-Mills algebra is the cubic algebra generated by $s+1$ elements $\theta^\lambda$ ($\lambda\in \{0,\dots,s\}$) with relations
\[
\theta^\lambda \theta^\mu\theta^\nu=\frac{1}{s}(g^{\lambda\mu}\theta^\nu+g^{\mu\nu}\theta^\lambda - 2g^{\lambda\nu} \theta^\mu) \gg
\]
where $\gg=g_{\alpha\beta} \theta^\alpha \theta^\beta$. These relations imply that $\gg \in \cala^!_2$ is central in $\cala^!$ and that one has $\cala^!_0=\mathbb K\bbbone \simeq \mathbb K$, $\cala^!_1=\oplus_\lambda \mathbb K \theta^\lambda \simeq \mathbb K^{s+1}$, $\cala^!_2=\oplus_{\mu\nu} \mathbb K \theta^\mu\theta^\nu \simeq \mathbb K^{(s+1)^2}$, $\cala^!_3=\oplus_\lambda \mathbb K \theta^\lambda \gg \simeq \mathbb K^{s+1}$, $\cala^!_4=\mathbb K \gg^2\simeq \mathbb K$ and $\cala^!_n=0$ for $n\geq 5$. From this, one obtains the description of \cite{ac-mdv:2002b} of the Koszul complex $\calk(\cala,\mathbb K)$ and the proof of the above theorem. It also follows that the bimodule resolution $\calk(\cala,\cala)\stackrel{\mu}{\rightarrow} \cala\rightarrow 0$ of $\cala$ reads
\begin{equation}
0\rightarrow \cala\otimes \cala \stackrel{\delta^\prime_3}{\rightarrow}\cala \otimes \mathbb K^{s+1}\otimes \cala \stackrel{\delta^\prime_2}{\rightarrow} \cala \otimes \mathbb K^{s+1} \otimes \cala \stackrel{\delta^\prime_1}{\rightarrow}\cala\otimes \cala \stackrel{\mu}{\rightarrow} \cala\rightarrow 0
\label{eq3.2}
\end{equation}
where the components $\delta^\prime_k$ of $\delta^\prime$ in the different degrees can be computed by using the description of $\calk(\cala,\mathbb K)=C_{2,0}$ given in Section 3 of 
\cite{ac-mdv:2002b} and are given by
\begin{equation}
\left\{
\begin{array}{l}
\delta^\prime_1(a\otimes e_\lambda \otimes b) = a\nabla_\lambda \otimes b - a \otimes \nabla_\lambda b\\
\\
\delta^\prime_2(a \otimes e_\lambda \otimes b) = 
(g^{\alpha\beta} \delta^\gamma_\lambda + g^{\beta\gamma} g^\alpha_\lambda - 2g^{\gamma\alpha}\delta^\beta_\lambda) \times \\
\hspace{1cm} \times (a\nabla_\alpha \nabla_\beta \otimes e_\gamma \otimes b + a\nabla_\alpha \otimes e_\gamma\otimes \nabla_\beta b+a\otimes e_\gamma \otimes \nabla_\alpha\nabla_\beta b)\\
\\
\delta^\prime_3(a\otimes b) = g^{\lambda\mu} (a\nabla_\mu\otimes e_\lambda \otimes b - a\otimes e_\lambda \otimes \nabla_\mu b)
\end{array}
\right.
\label{eq3.3}
\end{equation}
where $a,b\in \cala$, $e_\lambda$ ($\lambda=0,\dots, s$) is the canonical basis of $\mathbb K^{s+1}$ and $\nabla_\lambda$ are the corresponding generators of $\cala$.\\

Let $\calm$ be a bimodule over $\cala$. By using the above description of the Koszul resolution of $\cala$ one easily obtains the one of the small Hochschild complex $\cals(\cala,\calm)$ which reads 
\begin{equation}
0\rightarrow \calm \stackrel{\delta_3}{\rightarrow} \calm \otimes \mathbb K^{s+1} \stackrel{\delta_2}{\rightarrow} \calm \otimes \mathbb K^{s+1} \stackrel{\delta_1}{\rightarrow} \calm \rightarrow 0
\label{eq3.4}
\end{equation}
with differential $\delta$ given by
\begin{equation}
\left\{
\begin{array}{l}
\delta_1(m^\lambda\otimes e_\lambda)=m^\lambda \nabla_\lambda - \nabla_\lambda m^\lambda=[m^\lambda,\nabla_\lambda]\\
\\
\delta_2(m^\lambda \otimes e_\lambda)=\\
\hspace{0,5cm}=([\nabla_\mu,[\nabla^\mu,m^\lambda]]+[\nabla_\mu,[m^\mu,\nabla^\lambda]] +[m^\mu,[\nabla_\mu,\nabla^\lambda]])\otimes e_\lambda\\
\\
\delta_3(m) = g^{\lambda\mu}(m\nabla_\mu-\nabla_\mu m)\otimes e_\lambda=[m,\nabla^\lambda]\otimes e_\lambda
\end{array}
\right.
\label{eq3.5}
\end{equation}
with obvious notations. By using (\ref{eq3.5}) one easily verifies the duality (\ref{Pd}). For instance $H_3(\cala,\calm)$ is $\ker(\delta_3)$ which is given by the $m\in\calm$ such that $\nabla_\lambda m=m\nabla_\lambda$ for $\lambda=0,\dots, s$ that is such that $am=ma$, $\forall a\in \cala$, since $\cala$ is generated by the $\nabla_\lambda$ and it is well known that this coincides with $H^0(\cala,\calm)$. Similarily $m^\lambda\otimes e_\lambda$ is in $\ker(\delta_2)$ if and only if $\nabla_\lambda \mapsto D(\nabla_\lambda)=g_{\lambda\mu}m^\mu$ extends as a derivation $D$ of $\cala$ into $\calm$ ($D\in \Der(\cala,\calm)$) while $m^\lambda\otimes e_\lambda=\delta_3(m)$ means that this derivation is inner $D=ad(m)\in \Int(\cala,\calm)$ from which $H_2(\cala,\calm)$ identifies with $H^1(\cala,\calm)$, and so on.\\

Assume now that $\calm$ is graded in the sense that one has $\calm=\oplus_{n\in \mathbb Z}\calm_n$ with $\cala_k \calm_\ell \cala_m \subset \calm_{k+\ell+m}$. Then the small Hochschild complex $\cals(\cala, \calm)$ splits into subcomplexes $\cals(\cala,\calm)=\oplus_n \cals^{(n)}(\cala,\calm)$ where $\cals^{(n)}(\cala,\calm)$ is the subcomplex
\begin{equation}
0\rightarrow \calm_{n-4} \stackrel{\delta_3}{\rightarrow} \calm_{n-3} \otimes \mathbb  K^{s+1} \stackrel{\delta_2}{\rightarrow} \calm_{n-1}\otimes \mathbb K^{s+1}\stackrel{\delta_1}{\rightarrow} \calm_n \rightarrow 0
\label{eq3.6}
\end{equation}
of (\ref{eq3.4}). Assume furthermore that the homogeneous components $\calm_n$ are finite-dimensional vector spaces, i.e. $\dim (\calm_n)\in \mathbb N$. Then one has the following Euler-Poincar\'e formula
\begin{equation}
\begin{array}{l}
\dim(H^{(n)}_0)-\dim(H^{(n)}_1)+\dim(H^{(n)}_2)-\dim(H^{(n)}_3)=\\
\\
\dim(\calm_n)-(s+1)\dim(\calm_{n-1}) + (s+1)\dim(\calm_{n-3})-\dim(\calm_{n-4})
\end{array}
\label{eq3.7}
\end{equation}
for the homology $H^{(n)}$ of the chain complex $\cals^{(n)}(\cala,\calm)$.\\

In the case where $\calm=\cala$, it follows from the Koszulity of $\cala$ that the right hand side of (\ref{eq3.7}) vanishes for $n\not= 0$. Denoting as usual by $HH(\cala)$ the $\cala$-valued Hochschild homology of $\cala$ which is here the homology of $\cals(\cala,\cala)$, we denote by $HH^{(n)}(\cala)$ the homology of the subcomplex $\cals^{(n)}(\cala,\cala)$. Since $\cala_n=0$ for $n<0$, one has $HH^{(n)}_0(\cala)=0$ for $n<0$, $HH^{(n)}_1(\cala)=0$ for $n\leq 0$, $HH^{(n)}_2(\cala)=0$ for $n\leq 2$ and $HH^{(n)}_3(\cala)=0$ for $n\leq 3$. Furthermore one has
\begin{equation}
HH^{(0)}_0(\cala) = HH^{(4)}_3(\cala) = \mathbb K
\label{eq3.8}
\end{equation}
\begin{equation}
HH^{(1)}_0(\cala) = HH^{(1)}_1(\cala) = HH^{(3)}_2(\cala) = \mathbb K^{s+1}
\label{eq3.9}
\end{equation}
\begin{equation}
HH^{(2)}_0(\cala) = HH^{(2)}_1(\cala) = \mathbb K^{\frac{(s+1)(s+2)}{2}}
\label{eq3.10}
\end{equation}
and the Euler Poincar\'e formula reads here
\begin{equation}
\dim(HH^{(n)}_0(\cala))+\dim(HH^{(n)}_2(\cala)) = \dim (HH^{(n)}_1 (\cala)) + \dim(HH^{(n)}_3(\cala))
\label{eq3.11}
\end{equation}
for $n\geq 1$ which implies 
\begin{equation}
\dim(HH^{(3)}_0(\cala)) + (s+1) = \dim(HH^{(3)}_1(\cala))
\label{3.12}
\end{equation}
for $n=3$ while for $n=1$ and $n=2$ it is already contained in (\ref{eq3.9}) and (\ref{eq3.10}).\\

The complete description of the Hochschild homology and of the cyclic homology of the Yang-Mills algebra will be given in \cite{ac-mdv:2005}.

\section{The super Yang-Mills algebra}

As pointed out in the introduction, the Yang-Mills algebra is the universal enveloping algebra of a Lie algebra which is graded by giving degree 1 to the generators $\nabla_\lambda$ (see in 
\cite{ac-mdv:2002b}). Replacing the Lie bracket by a super Lie bracket, that is replacing in the Yang-Mills equations (\ref{YM}) the commutator by the anticommutator whenever the 2 elements are of odd degrees, one obtains a super version $\tilde \cala$ of the Yang-Mills algebra $\cala$. In other words one defines the {\sl super Yang-Mills algebra} to be the cubic algebra $\tilde \cala$ generated $s+1$ elements $S_\lambda$ ($\lambda\in \{0,\dots, s\}$) with the relations
\begin{equation}
g^{\lambda\mu}[S_\lambda,\{S_\mu,S_\nu\}]=0,\>\>\> \nu\in \{0,\dots,s\}
\label{SYM}
\end{equation}
that is $\tilde \cala=A(\tilde E,\tilde R)$ with $\tilde E=\oplus_\lambda \mathbb KS_\lambda$ and $\tilde R\subset \tilde E^{\otimes^3}$ given by 
\begin{equation}
\tilde R=\sum_\rho \mathbb K(g^{\rho\lambda} g^{\mu\nu}-g^{\nu\rho} g^{\lambda\mu})S_\lambda \otimes S_\mu \otimes S_\nu
\label{rSYM}
\end{equation}
Relations (\ref{SYM}) can be equivalently written as
\begin{equation}
[g^{\lambda\mu} S_\lambda S_\mu,S_\nu]=0, \>\>\> \nu\in \{0,\dots,s\}
\label{ZSYM}
\end{equation}
which mean that $g^{\lambda\mu}S_\lambda S_\mu\in \tilde \cala_2$ is central in $\tilde\cala$.\\

It is easy to verify that the dual algebra $\tilde \cala^!=A(\tilde E^\ast, \tilde R^\perp)$ is the cubic algebra generated by $s+1$ elements $\xi^\lambda$ ($\lambda\in \{0,\dots,s\}$) with the relations
\[
\xi^\lambda \xi^\mu\xi^\nu=-\frac{1}{s} (g^{\lambda\mu}\xi^\nu-g^{\mu\nu}\xi^\lambda)\gg
\]
where $\gg=g_{\alpha\beta}\xi^\alpha \xi^\beta$. These relations imply that $\gg \xi^\nu+\xi^\nu \gg=0$, i.e. 
\begin{equation}
\{g_{\lambda\mu} \xi^\lambda \xi^\mu,\xi^\nu\}=0,\>\>\> \nu\in \{0,\dots, s\}
\label{AZ}
\end{equation}
and that one has $\tilde \cala_0^!=\mathbb K\bbbone \simeq \mathbb K$, $\tilde \cala^!_1=\oplus_\lambda \mathbb K \xi^\lambda \simeq \mathbb K^{s+1}$, $\tilde \cala^!_2=\oplus_{\mu\nu}\mathbb K \xi^\mu \xi^\nu \simeq \mathbb K^{(s+1)^2}$, $\tilde\cala^!_3=\oplus_\lambda \mathbb K \xi^\lambda \gg \simeq \mathbb K^{s+1}$, $\tilde\cala^!_4=\mathbb K \gg^2 \simeq \mathbb K$ and $\tilde\cala^!_n=0$ for $n\geq 5$.\\

The Koszul complex $\calk(\tilde\cala,\mathbb K)$ of $\tilde\cala$ then reads 
\[
0\rightarrow \tilde \cala\stackrel{S^t}{\rightarrow} \tilde\cala^{s+1}\stackrel{N}{\rightarrow} \tilde\cala^{s+1}\stackrel{S}{\rightarrow} \tilde \cala \rightarrow 0
\]
where $S$ means right multiplication by the column with components $S_\lambda$, $S^t$ means right multiplication by the row with components $S_\lambda$ and $N$ means right multiplication (matrix product) by the matrix with components
\[
N^{\mu\nu}=(g^{\mu\nu} g^{\alpha\beta}-g^{\mu\alpha}g^{\nu\beta})S_\alpha S_\beta
\]
with $\lambda, \mu, \nu\in \{0,\dots, s\}$. One has the following result.
\begin{theorem}
The cubic super Yang-Mills algebra $\tilde \cala$ is Koszul of global dimension 3 and is Gorenstein.
\end{theorem}

\noindent \underbar{Proof}. By the very definition of $\tilde\cala$ by generators and relations, the sequence
\[
\tilde\cala^{s+1}\stackrel{N}{\rightarrow} \tilde\cala^{s+1}\stackrel{S}{\rightarrow} \tilde \cala \stackrel{\varepsilon}{\rightarrow} \mathbb K \rightarrow 0
\]
is exact. On the other hand it is easy to see that the mapping $\tilde\cala \stackrel{S^t}{\rightarrow} \tilde\cala^{s+1}$ is injective and that the sequence
\[
0\rightarrow \tilde \cala \stackrel{S^t}{\rightarrow}\tilde\cala^{s+1} \stackrel{N}{\rightarrow} \tilde\cala^{s+1} \stackrel{S}{\rightarrow} \tilde \cala \stackrel{\varepsilon}{\rightarrow}\mathbb K \rightarrow 0
\]
is exact which implies that $\tilde\cala$ is Koszul of global dimension 3. The Gorenstein property follows from the symmetry by transposition. $\square$\\

The situation is completely similar to the Yang-Mills case, in particular $\tilde\cala$ has Hochschild dimension 3 and, by applying a result of 
\cite{mdv-pop:2002}, $\tilde\cala$ has the same Poincar\'e series as $\cala$ i.e. one has the formula
\begin{equation}
\sum_{n\in \mathbb N} \dim(\tilde\cala_n) t^n=\frac{1}{(1-t^2)(1-(s+1)t+t^2)}
\label{psa}
\end{equation}
which, as will be shown elsewhere, can be interpreted in terms of the quantum group of the bilinear form $(g_{\mu\nu})$ \cite{mdv-lau:1990}  by noting the invariance of Relations (\ref{ZSYM}) by this quantum group. For $s=1$ the Yang-Mills algebra and the super Yang-Mills algebra are particular cubic Artin-Schelter algebras \cite{art-sch:1987} whereas for $s\geq2$ these algebras have exponential growth as follows from Formula 
 (\ref{psa}).

\section{The super self-duality algebra}

There are natural quotients $\calb$ of $\cala$ and $\tilde\calb$ of $\tilde\cala$ which are connected with parastatistics and which have been investigated in \cite{mdv-pop:2002}. The {\sl parafermionic algebra} $\calb$ is the cubic algebra generated by elements $\nabla_\lambda$ ($\lambda\in \{0,\dots, s\}$) with relations
\[
[\nabla_\lambda,[\nabla_\mu, \nabla_\nu]]=0
\]
for any $\lambda, \mu,\nu\in \{0,\dots,s\}$, while the {\sl parabosonic algebra} $\tilde\calb$ is the cubic algebra generated by elements $S_\lambda$ ($\lambda\in \{0,\dots,s\}$) with relations
\[
[S_\lambda,\{S_\mu,S_\nu\}]=0
\]
for any $\lambda, \mu,\nu\in \{0,\dots,s\}$. In contrast to the Yang-Mills and  the super Yang-Mills algebras $\cala$ and $\tilde\cala$ which have exponential growth whenever $s\geq 2$, these algebras $\calb$ and $\tilde\calb$ have polynomial growth with Poincar\'e series given by
\[
\sum_n \dim(\calb_n) t^n=\sum_n \dim(\tilde\calb_n) t^n=\left(\frac{1}{1-t}\right )^{s+1}\left( \frac{1}{1-t^2}\right)^{\frac{s(s+1)}{2}}
\]
but they are not Koszul for $s\geq2$, \cite{mdv-pop:2002}.\\

In a sense, the algebra $\calb$ can be considered to be  somehow trivial from the point of view of the classical Yang-Mills equations in dimension $s+1\geq 3$ although the algebras $\calb$ and $\tilde \calb$ are quite interesting for other purposes \cite{mdv-pop:2002}. It turns out that in dimension $s+1=4$ with $g_{\mu\nu}=\delta_{\mu\nu}$ (Euclidean case), the Yang-Mills algebra $\cala$ has non trivial quotients $\cala^{(+)}$ and $\cala^{(-)}$ which are quadratic algebras refered to as the {\sl self-duality algebra} and the {\sl anti-self-duality algebra} respectively \cite{ac-mdv:2002b}. Let $\varepsilon=\pm$, the algebra $\cala^{(\varepsilon)}$ is the quadratic algebra generated by the elements $\nabla_\lambda$ ($\lambda\in \{0,1,2,3,\}$) with relations
\[
[\nabla_0,\nabla_k]=\varepsilon[\nabla_\ell,\nabla_m]
\]
for any cyclic permutation $(k,\ell,m)$ of (1,2,3). One passes from $\cala^{(-)}$ to $\cala^{(+)}$ by changing the orientation of $\mathbb K^4$ so one can restrict attention to the self-duality algebra $\cala^{(+)}$. This algebra has been studied in \cite{ac-mdv:2002b}
where it was shown in particular that it is Koszul of global dimension 2. For further details on this algebra and on the Yang-Mills algebra, we refer to \cite{ac-mdv:2002b} and to the forthcoming paper \cite{ac-mdv:2005}. Our aim now in this section is to define and study the super version of the self-duality algebra.\\

Let $\varepsilon=+$ or $-$ and define $\tilde\cala^{(\varepsilon)}$ to be the quadratic algebra generated by the elements $S_0, S_1, S_2, S_3$ with relations
\begin{equation}
i\{S_0,S_k\}=\varepsilon [S_\ell, S_m]
\label{essd}
\end{equation}
for any cyclic permutation $(k,\ell,m)$ of (1,2,3). One has the following.
\begin{lemma}
Relations (\ref{essd}) imply that one has
\[
\big [\sum^3_{\mu=0}(S_\mu)^2,S_\lambda\big]=0
\]
for any $\lambda\in \{0,1,2,3\}$. In other words, $\tilde\cala^{(+)}$ and $\tilde\cala^{(-)}$ are quotients of the super Yang-Mills algebra $\tilde\cala$ for s+1=4 and $g_{\mu\nu}=\delta_{\mu\nu}$.
\end{lemma}
The proof which is a straightforward verification makes use of the Jacobi identity (see also in \cite{skl:1982}). Thus $\tilde\cala^{(+)}$ and $\tilde\cala^{(-)}$ play the same role with respect to $\tilde\cala$ as $\cala^{(+)}$ and $\cala^{(-)}$ with respect to $\cala$. Accordingly they will be respectively called the {\sl super self-duality algebra} and the {\sl super anti-self-duality algebra}. Again $\tilde\cala^{(+)}$ and $\tilde\cala^{(-)}$ are exchanged by changing the orientation of $\mathbb K^4$ and we shall restrict attention to the super self-duality algebra in the following, i.e. to the quadratic algebra $\tilde\cala^{(+)}$ generated by $S_0,S_1,S_2, S_3$ with relations
\begin{equation}
i\{S_0,S_k\}=[S_\ell,S_m]
\label{ssd}
\end{equation}
for any cyclic permutation $(k,\ell,m)$ of (1,2,3). One has the following result.

\begin{theorem}
The quadratic super self-duality algebra $\tilde\cala^{(+)}$ is a Koszul algebra of global dimension 2.
\end{theorem}
\noindent \underbar{Proof}. One verifies that the dual quadratic algebra $\tilde\cala^{(+)!}$ is generated by elements $\xi^0,\xi^1,\xi^2,\xi^3$ with relations $(\xi^\lambda)^2=0,$ for $\lambda=0,1,2,3$ and $\xi^\ell\xi^m=-\xi^m\xi^\ell=i\xi^0\xi^k=i\xi^k\xi^0$ for any cyclic permutation $(k,\ell,m)$ of (1,2,3). So one has $\tilde\cala^{(+)!}_0=\mathbb K\bbbone \simeq \mathbb K$, $\tilde \cala^{(+)!}_1=\oplus_\lambda \mathbb K \xi^\lambda\simeq \mathbb K^4$, $\tilde\cala^{(+)!}_2=\oplus_k\mathbb K\xi^0\xi^k\simeq \mathbb K^3$ and $\tilde\cala^{(+)!}_n=0$ for $n\geq 3$ since the above relations imply $\xi^\lambda\xi^\mu\xi^\nu=0$ for any $\lambda,\mu,\nu\in\{0,1,2,3\}$. The Koszul complex $K(\tilde\cala^{(+)})=\calk(\tilde\cala^{(+)},\mathbb K)$ (quadratic case) then reads
\[
0\rightarrow \tilde\cala^{(+)^3}\stackrel{D}{\rightarrow} \tilde\cala^{(+)^4}\stackrel{S}{\rightarrow} \tilde\cala^{(+)}\rightarrow 0
\]
where $S$ means right matrix product with the column with components $S_\lambda$ ($\lambda\in \{0,1,2,3\}$) and $D$ means right matrix product with 

\begin{equation}
D= \left (\begin{array}{cccc}
iS_1 & iS_0 & S_3 & -S_2\\
iS_2 & -S_3 & iS_0 & S_1\\
iS_3 & S_2 & -S_1 & iS_0
\end{array}
\right )
\label{D}
\end{equation}
It follows from the definition of $\tilde\cala^{(+)}$ by generators and relations that the sequence
\[
\tilde\cala^{(+)^3}\stackrel{D}{\rightarrow} \tilde\cala^{(+)^4}\stackrel{S}{\rightarrow} \tilde\cala^{(+)}\stackrel{\varepsilon}{\rightarrow} \mathbb K \rightarrow 0
\]
is exact. On the other hand one shows easily that the mapping $\tilde\cala^{(+)^3}\stackrel{D}{\rightarrow} \tilde\cala^{(+)^4}$ is injective so finally the sequence 
\[
0\rightarrow \tilde\cala^{(+)3} \stackrel{D}{\rightarrow} \tilde\cala^{(+)^4}\stackrel{S}{\rightarrow} \tilde\cala^{(+)}\stackrel{\varepsilon}{\rightarrow} \mathbb K \rightarrow 0
\]
is exact which implies the result. $\square$\\

This theorem implies that the super self-duality algebra $\tilde\cala^{(+)}$ has Hochschild dimension 2 and that its Poincar\'e series is given by
\[
P_{\tilde\cala^{(+)}}(t)=\frac{1}{(1-t)(1-3t)}
\]
in view of the structure of its dual $\tilde\cala^{(+)!}$ described in the proof. Thus everything is similar to the case of the self-duality algebra $\cala^{(+)}$.\\

Let us recall that the {\sl Sklyanin algebra}, in the presentation given by Sklyanin \cite{skl:1982}, is the quadratic algebra $\cals(\alpha_1,\alpha_2,\alpha_3)$ generated by 4 elements $S_0,S_1,S_2,S_3$ with relations
\[
i\{S_0,S_k\}=[S_\ell,S_m]
\]
\[
[S_0,S_k]=i\frac{\alpha_\ell-\alpha_m}{\alpha_k} \{S_\ell, S_m\}
\]
for any cycic permutation $(k,\ell, m)$ of (1,2,3). One sees that the relations of the super self-duality algebra $\tilde\cala^{(+)}$ are the relations of the Sklyanin algebra which are independent from the parameters $\alpha_k$. Thus one has a sequence of surjective homomorphisms of connected graded algebra
\[
\tilde\cala \rightarrow \tilde\cala^{(+)} \rightarrow \cals (\alpha_1,\alpha_2,\alpha_3)
\]
On the other hand for {\sl generic values of the parameters} the Sklyanin algebra is Koszul Gorenstein of global dimension 4 
\cite{smi-sta:1992} with the same Poincar\'e series as the polynomial algebra $\mathbb K [X_0,X_1,X_2,X_3]$ and corresponds to the natural ambiant noncommutative 4-dimensional Euclidean space containing  the noncommutative 3-spheres described in \cite{ac-mdv:2002a}, 
\cite{ac-mdv:2003} (their ``homogeneisation"). This gives a very surprising connection between the present study and our noncommutative 3-spheres for generic values of the parameters. It is worth noticing here that in the analysis of 
\cite{ac-mdv:2003} several bridges between noncommutative differential geometry in the sense of 
\cite{ac:1986a}, \cite{ac:1994} and noncommutative algebraic geometry have been established.

\section{Deformations}

The aim of this section is to study deformations  of the Yang-Mills algebra and of the super Yang-Mills algebra. We use the notations of Sections 3 and 4.\\

Let the dimension $s+1\geq 2$ and the pseudo metric $g_{\lambda\mu}$ be fixed and let $\zig \in  P_1(\mathbb K)$ have homogeneous coordinates $\zig_0,\zig_1\in \mathbb K$. Define $\cala(\zig)$ to be the cubic algebra generated by $s+1$ elements $\nabla_\lambda$ ($\lambda\in \{0,\dots, s\}$) with relations
\[
(\zig_1(g^{\rho\lambda} g^{\mu\nu}+ g^{\nu \rho} g^{\lambda\mu})-2\zig_0 g^{\rho \mu} g^{\lambda\nu})\nabla_\lambda \nabla_\mu\nabla_\nu=0
\]
for $\rho\in \{0,\dots, s\}$. The Yang-Mills algebra corresponds to the element $\zig^{YM}$ of $P_1(\mathbb K)$ with homogeneous coordinates $\zig_0=\zig_1$. Let $\zig^{sing}$ be the element of $P_1(\mathbb K)$ with homogeneous coordinates $\zig_0=\frac{s+2}{2}\zig_1$; one has the following result.

\begin{theorem}
For $\zig\not= \zig^{sing}$ the cubic algebra $\cala(\zig)$ is Koszul of global dimension 3 and is Gorenstein.
\end{theorem}
\noindent \underbar{Proof}. The dual algebra $\cala(\zig)^!$ is the cubic algebra generated by elements $\theta^\lambda$ with relations
\begin{equation}
\theta^\lambda\theta^\mu\theta^\nu=\frac{1}{(s+2)\zig_1-2\zig_0} (\zig_1(g^{\lambda\mu}\theta^\nu+g^{\mu\nu}\theta^\lambda)-2\zig_0 g^{\lambda\nu} \theta^\mu)\gg
\label{DDYM}
\end{equation}
for $\lambda, \mu,\nu\in \{0,\dots,s\}$ with $\gg=g_{\alpha\beta} \theta^\alpha \theta^\beta$. This again implies that $\gg$ is in the center and that one has $\cala^!_0=\mathbb K \bbbone \simeq \mathbb K$, $\cala^!_1=\oplus_\lambda \mathbb K \theta^\lambda \simeq \mathbb K^{s+1}$, $\cala^!_2=\oplus_{\lambda,\mu} \mathbb K \theta^\lambda\theta^\mu\simeq \mathbb K^{(s+1)^2}$, $\cala^!_3=\oplus_\lambda \mathbb K \theta^\lambda \gg \simeq \mathbb K^{s+1}$, $\cala^!_4=\mathbb K\gg^2 \simeq \mathbb K$ while $\cala^!_n=0$ for $n\geq 5$, where we have set $\cala^!_n=\cala(\zig)^!_n$. Setting $\cala=\cala(\zig)$, the Koszul complex  $\calk(\cala(\zig),\mathbb K)$ of $\cala(\zig)$ reads
\[
0 \rightarrow \cala \stackrel{\nabla^t}{\rightarrow} \cala^{s+1} \stackrel{M}{\rightarrow} \cala^{s+1} \stackrel{\nabla}{\rightarrow} \cala \rightarrow 0
\]
with the same conventions as before and $M$ with components
\[
M^{\mu\nu}=\frac{1}{(s+2)\zeta_1-2\zeta_0} (\zeta_1(g^{\mu\nu}g^{\alpha\beta}+g^{\mu\alpha}g^{\nu\beta})-2\zeta_0 g^{\mu\beta}g^{\nu\alpha})\nabla_\alpha\nabla_\beta
\]
$\mu, \nu, \in\{ 0,\dots,s\}$. The theorem follows then by the same arguments as before, using in particular the symmetry by transposition for the Gorenstein property. $\square$\\

It follows that $\cala(\zig)$ has Hochschild dimension 3 and the same Poincar\'e series as the Yang-Mills algebra for $\zig\not= \zig^{sing}$.\\

\noindent \underbar{Remark}. One can show that the cubic algebra generated by elements $\nabla_\lambda$ with relations
\[
(\zeta_1 g^{\rho\lambda}g^{\mu\nu}+\zeta_2 g^{\nu\rho} g^{\lambda\mu}-2 \zeta_0 g^{\rho\mu} g^{\lambda\nu}) \nabla_\lambda \nabla_\mu \nabla_\nu=0
\]
cannot be Koszul and Gorenstein if $\zeta_1\not= \zeta_2$ and $\zeta_0\not= 0$ or if $(\zeta_1)^2\not= (\zeta_2)^2$.\\

Let now $(B_{\lambda\mu})\in M_{s+1}(\mathbb K)$ be an arbitrary invertible $(s+1)\times (s+1)$-matrix with inverse $(B^{\lambda\mu})$, i.e. $B_{\lambda\mu}B^{\mu\nu}=\delta^\nu_\lambda$, and let $\varepsilon=+$ or $-$. We define $\fraca(B,\varepsilon)$ to be the cubic algebra generated by $s+1$ elements $E_\lambda$ with relations
\begin{equation}
(B^{\rho\lambda} B^{\mu\nu}+\varepsilon B^{\lambda\mu}B^{\nu\rho})E_\lambda E_\mu E_\nu =0
\label{BR}
\end{equation}
for $\rho\in \{0,\dots,s\}$. Notice that $B$ is not assumed to be symmetric. If $B_{\lambda\mu}=g_{\lambda\mu}$ and $\varepsilon=-$ then $\fraca(g,-)$ is the super Yang-Mills algebra $\tilde\cala$ $(E_\lambda\mapsto S_\lambda)$ while if $B_{\lambda\mu}=g_{\lambda\mu}$ and $\varepsilon=+$ then $\fraca(g,+)$  is $\cala(\zeta^0)$  $(E_\lambda\mapsto \nabla_\lambda)$ where $\zeta^0$ has homogeneous coordinates $\zeta_1\not=0$ and $\zeta_0=0$. Thus $\fraca(B,+)$ and $\fraca(B,-)$ belong to deformations of the Yang-Mills and of the super Yang-Mills algebra respectively.
\begin{theorem}
Assume that $1+\varepsilon B^{\rho\lambda} B^{\mu\nu} B_{\mu\lambda}B_{\rho\nu}\not= 0$, then $\fraca(B,\varepsilon)$ is Koszul of global dimension 3 and is Gorenstein.
\end{theorem}

\noindent\underbar{Proof}. The Koszul complex $\calk(\fraca(B,\varepsilon),\mathbb K)$ can be put in the form 
\[
0\rightarrow \fraca \stackrel{E^t}{\rightarrow} \fraca^{s+1}\stackrel{L}{\rightarrow} \fraca^{s+1}\stackrel{E}{\rightarrow} \fraca \rightarrow 0
\]
where $\fraca=\fraca(B,\varepsilon)$ and with the previous conventions, the matrix $L$ being given by
\begin{equation}
L^{\mu\nu}=(B^{\mu\alpha} B^{\beta\nu}+\varepsilon B^{\nu\mu} B^{\alpha\beta}) E_\alpha E_\beta
\label{mB}
\end{equation}
for $\mu,\nu\in \{0,\dots,s\}$. The arrow $\fraca\stackrel{E^t}{\rightarrow} \fraca^{s+1}$ is always injective and the exactness of $\fraca\stackrel{E^t}{\rightarrow}\fraca^{s+1} \stackrel{L}{\rightarrow} \fraca^{s+1}$ follows from the condition $1+\varepsilon B^{\rho\lambda}B^{\mu\nu} B_{\mu\lambda} B_{\rho\nu}\not= 0$. On the other hand,  by definition of $\cala$ by generators and relations,  the sequence $\fraca^{s+1}\stackrel{L}{\rightarrow} \fraca^{s+1} \stackrel{E}{\rightarrow} \fraca \stackrel{\varepsilon}{\rightarrow} \mathbb K \rightarrow 0$ is exact. This shows that $\fraca$ is Koszul of global dimension 3. The Gorenstein property follows from (see also in \cite{art-sch:1987})
\[
B^{\rho\lambda}B^{\mu\nu} + \varepsilon B^{\nu\rho} B^{\lambda\mu}=\varepsilon (B^{\nu\rho}B^{\lambda\mu}+\varepsilon B^{\mu\nu}B^{\rho\lambda})
\]
for $\rho, \lambda, \mu,\nu \in \{0,\dots, s\}$. $\square$\\

\noindent \underbar{Remark}. It is worth noticing here in connection with the analysis of \cite{ber-mar:2003} that for all the deformations of the Yang-Mills algebra (resp. the super Yang-Mills algebra) considered here which are cubic Koszul Gorenstein algebras of global dimension 3, the dual cubic algebras are Frobenius algebras with structure automorphism equal to the identity (resp. $(-1)^{\mbox{degree}} \times$ identity).

\bibliographystyle{plain}
\bibliography{BibMich,BibExt}

\end{document}